\documentclass{article}
\usepackage{spconf,amsmath,graphicx}
\usepackage{dsfont}
\usepackage{algorithmic}
\usepackage{amsfonts,amssymb}
\usepackage[]{subfigure}
\usepackage{xcolor}

\usepackage{balance}
\usepackage[font=small,skip=0pt]{caption}

\usepackage{color, soul}

\DeclareMathOperator*{\argminA}{arg\,min}

\title{\vspace{-6mm}Intelligent Environments based on Ultra-Massive MIMO Platforms for Wireless Communication in Millimeter Wave and Terahertz Bands}
\name{Shuai~Nie$^{1}$, Josep~M.~Jornet$^{2}$, and~Ian~F.~Akyildiz$^{1}$\thanks{This work was supported by the U.S. National Science Foundation under Grant No. ECCS-1608579 and in part by Alexander von Humboldt Foundation through Ian Akyildiz's Humboldt Research Prize in Germany.}}
\address{\small$^1$Broadband Wireless Networking Lab, School of Electrical and Computer Engineering,
\small{Georgia Institute of Technology, USA}\\
		\small$^2$Department of Electrical Engineering,
		\small{University at Buffalo, The State University of New York, USA}}
\begin{document}
%
\maketitle
\begin{abstract}
Millimeter-wave (30-300 GHz) and Terahertz-band communications (0.3-10 THz) are envisioned as key wireless technologies to satisfy the demand for Terabit-per-second (Tbps) links in the 5G and beyond eras. The very large available bandwidth in this ultra-broadband frequency range comes at the cost of a very high propagation loss, which combined with the low power of mm-wave and THz-band transceivers limits the communication distance and data-rates. In this paper, the concept of intelligent communication environments enabled by Ultra-Massive MIMO platforms is proposed to increase the communication distance and data-rates at mm-wave and THz-band frequencies. An end-to-end physical model is developed by taking into account the capabilities of novel intelligent plasmonic antenna arrays which can operate in transmission, reception, reflection and waveguiding, as well as the peculiarities of the mm-wave and THz-band multi-path channel. Based on the developed model, extensive quantitative results for different scenarios are provided to illustrate the performance improvements in terms of both achievable distance and data-rate in Ultra-Massive MIMO environments.
\end{abstract}
%
\section{Introduction}
\label{sec:intro}
In recent decades, a dramatic revolution in wireless communication networks has been witnessed. Among others, approximately 20.4~billion wireless devices are expected to be connected by~2020 due to the rise of the Internet of Things paradigm. In parallel to the growth of the number of wireless devices, services, and applications, there has been an increasing demand for higher speed wireless communications. To meet this demand, new spectral bands and advanced communication techniques are needed.  The research in millimeter wave (mm-wave, 30--300~GHz) and Terahertz (THz) frequencies (300~GHz -- 10~THz) has attracted great attention in recent years owing to their intrinsically abundant spectrum resources, ranging from 10~GHz in the mm-wave band to several windows each with a few hundred GHz consecutive spectrum in the THz band. If properly utilized, such massive spectral resources can lead to drastic improvements in terms of individual user data-rate and multi-user throughput.

Nevertheless, the major challenge at mm-wave and THz-band frequencies is the limited communication distance because of the remarkably high path loss inherent to small wavelengths and the limited transmission power of mm-wave and THz-band transceivers~\cite{akyildiz2018combating}. To overcome this problem, high-gain directional antenna systems are needed. Besides, reconfigurable antenna arrays can be utilized to implement MIMO and Massive MIMO communication systems. When increasing the communication frequency, antennas become smaller and, thus, a higher number of antennas can be integrated within the same footprint. However, simply increasing the number of antennas is not sufficient to overcome the much higher path loss as we move towards the THz band and to meet the demands and expectations of beyond 5G systems.


 \begin{figure*}[h]
	\centering
	\includegraphics[width=0.95\textwidth]{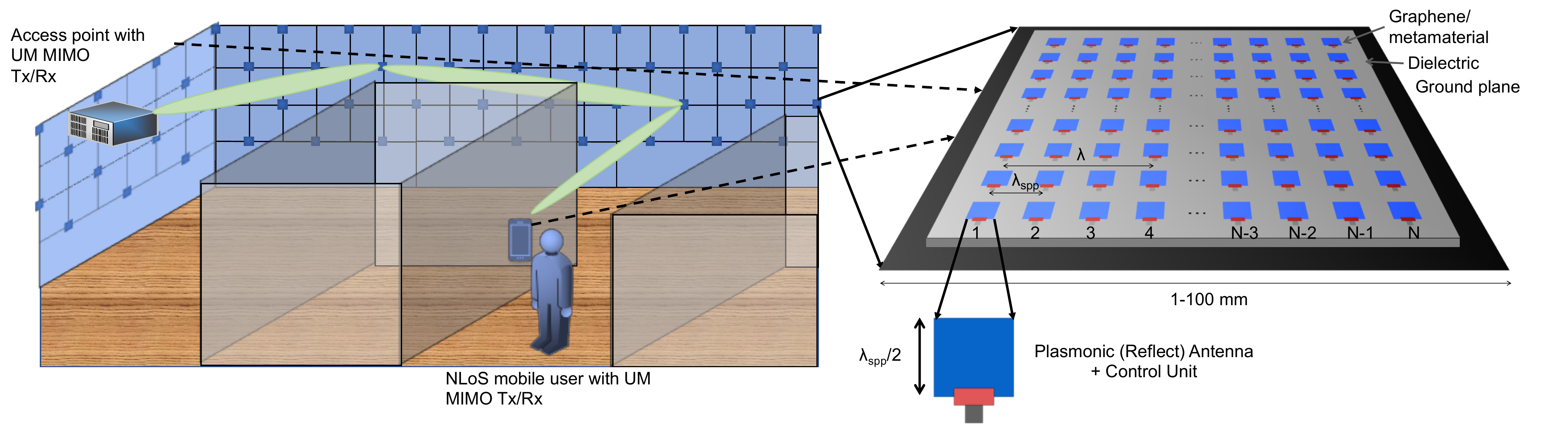} 
	\caption{A conceptual design of intelligent environments and the plasmonic antenna (reflect) array.}
	\label{fig:concept}
	\vspace{-4mm}
\end{figure*}

In this paper, we propose the concept of intelligent communication environments enabled by Ultra-Massive MIMO (UM MIMO) platforms as a way to increase the communication distance and data-rates at mm-wave and THz-band frequencies. The UM MIMO platforms consist of reconfigurable plasmonic antenna arrays both at the transmitting and receiving nodes, operating as \textit{plasmonic transmit-receive arrays}, and in the transmission environment, in the form of \textit{plasmonic reflect-arrays} and able to operate in different modes, including transmission, reception, reflection, and waveguiding. Plasmonic antennas leverage the physics of Surface Plasmon Polariton (SPP) waves to efficiently radiate at the target resonant frequency while being much smaller than the corresponding wavelength~\cite{jornet2013graphene}. This particular property allows them to be integrated in very dense arrays, beyond traditional antenna arrays, and enables the precise radiation and propagation control of EM waves with sub-wavelength resolution.

 Comparing to prior works, in~\cite{akyildiz2016realizing}, we introduced for the first time the concept of UM MIMO at the transmitter and the receiver. In~\cite{liaskos2018new, Liaskos2018using}, the use of programmable metasurfaces or \emph{HyperSurfaces} was proposed as a way to engineer the propagation of electromagnetic signals in the channel, with more control than with traditional programmable reflect-arrays~\cite{tan2018enabling}. Here, this is the first work in which full-wave control with plasmonic arrays at the transmitter, through the channel, and at the receiver is proposed. We analytically model and numerically estimate the performance of intelligent communication environments (Fig.~\ref{fig:concept}) based on UM MIMO systems at the transmitter, the receiver and deployed through the communication channel in a synergistic operation manner to overcome the main challenge at mm-wave and THz frequencies.

\vspace{-2mm}

\section{Design of UM MIMO Platforms-Based Intelligent Environments}
\label{sec:hardware}
%

UM MIMO platforms enable the creation of intelligent communication environments in both indoor and outdoor scenarios. With the derivations readily extendable to the outdoor case, here we focus our analysis in the indoor case. The intelligent environments consist of two major parts: the plasmonic transmit-receive arrays at the nodes and the plasmonic reflectarray systems in the propagation environment. Plasmonic reflectarrays can be embedded or applied to surfaces of indoor objects (e.g., walls and ceilings) like adhesive foil papers with low energy cost and allow signal transmissions through reflections on the plasmonic layer (i.e., the top layer) or waveguiding on the waveguiding layer (i.e., the bottom layer). The control layer (i.e., implemented in the middle layer) estimates the channel, coordinates with the transmitting and receiving arrays at the nodes, and assigns the operation modes to individual or groups of plasmonic units. The UM MIMO platforms are powered by batteries in mobile transceivers and AC power supply for wall-covering reflectarrays. 

\vspace{-2mm}
\subsection{Plasmonic Transmit--Receive Arrays}


 The physics of plasmonics can be utilized to enable antenna arrays with much denser elements and go beyond the conventional $\lambda/2$ sampling of space towards more precise space and frequency beamforming. On that basis, we have demonstrated that, at THz frequencies, graphene can be used to build nano-transceivers and nano-antennas with maximum dimension $\lambda/20$, allowing them to be densely integrated in very small footprints (1024 elements in less than 1 $\textrm{mm}^2$), as shown in Fig.~\ref{fig:concept}~\cite{jornet2013graphene,jornet2014graphene}. However, graphene does not perform well at mm-wave bands. Instead, we consider to incorporate metasurfaces into the plasmonic transmit-receive arrays. Metasurfaces, the 2D representation of metamaterials  which are a type of engineered material designed to exhibit EM properties not commonly found in nature, have been well studied from the perspective of material science and technology~\cite{lockyear2009microwave}. The understanding of the benefits of utilizing them in transmission and reception from the communication perspective is missing and motivates this work. 
 
\begin{figure*}[!h]
\centering
        \includegraphics[width=0.55\textwidth]{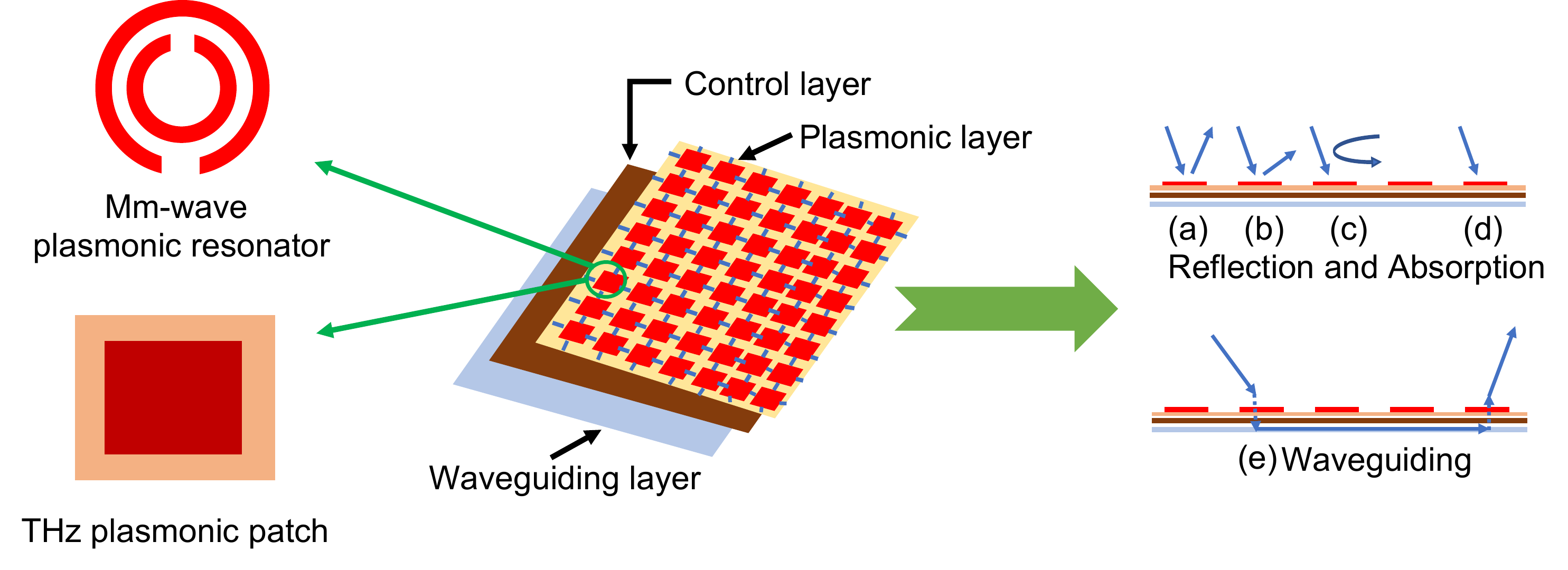} 
        \caption{Conceptual design of a plasmonic reflectarray able to unconventionally manipulate EM waves, including (a) specular reflection; (b) controlled reflection; (c) reflection with converted polarization; (d) absorption; and (e) signal waveguiding. }
        \label{fig:meta_description}
        	\vspace{-4mm}
\end{figure*}

\vspace{-2mm}
\subsection{Plasmonic Reflectarray System in the Intelligent Environments}
The plasmonic reflectarrays can be deployed freely in the 3D environment, with a size ranging from 1~mm$^2$ to 100~mm$^2$ depending on the operating frequency (mm-wave/THz band) equipped with hundreds or thousands of plasmonic antenna elements. Owing to the sub-wavelength size of their elements, the plasmonic reflectarrays are able to reflect signals in non-conventional ways, which include controlled reflections in non-specular directions as well as reflections with polarization conversion~\cite{zhu2013linear}, as illustrated in Fig.~\ref{fig:meta_description}.  These operations can be achieved by properly tuning the reflection coefficient $\Gamma$ of the meta-atoms. To be more specific, suppose a pair of incident and reflected signal has the following wave vectors,
%
%
\vspace{-2mm}
%
\allowdisplaybreaks
\begin{align}
\mathbf{E}^{\mathrm{in}} &= \left(E_x\hat{x} + E_y\hat{y}\right)e^{i(-\omega z/c + \omega t)},\\ \label{eq:wave_in}
\mathbf{E}^{\mathrm{ref}} &= \Gamma^b \left(E_x\hat{x} + E_y\hat{y}\right)e^{i(\omega z/c + \omega t)},
\end{align} where $\omega$ is the angular frequency. The reflection coefficient $\Gamma$ can be written as $\Gamma = \mathbf{E}^{\mathrm{ref}} / \mathbf{E}^{\mathrm{in}} \nonumber
 =\Gamma^{b} e^{i(2\omega z/c)} = \Gamma^b e^{i\Phi}$,  where $\Gamma^b$ is the reflection coefficient at the boundary of two materials.
We can further express the phase component of the reflection coefficient as a function between the angles of incidence and reflection, $e^{i\Phi} = e^{i(\sin\theta_{\mathrm{in}} - \sin\theta_{\mathrm{ref}})\omega/c z}, \frac{d\Phi(z)}{dz} = \omega z/c(\sin\theta_{\mathrm{in}} - \sin\theta_{\mathrm{ref}})$. Hence by controlling the phase component, the reflection signals can be tuned to a desired direction. 


The polarization conversion can be achieved by altering the reflection coefficients along two orthogonal directions in anisotropic meta-atoms~\cite{hao2007manipulating}, where the wave vector of reflected wave is expressed as $\mathbf{E}^{\mathrm{ref}} = (\Gamma_x E_x\hat{x} + \Gamma_y E_y\hat{y})e^{i(\omega z/c + \omega t)}$.
One step further, the absorption of incident EM waves can be maximized in metasurfaces with proper size and thickness selections of the meta-atoms.  According to conservation of energy, the absorptivity $\beta$ can be maximized when the reflectivity $R$ is minimized, where we have $\beta = 1-R = 1-\left |{}\frac{\mu_r - n}{\mu_r + n}  \right |^2, n = \sqrt{\mu_r\varepsilon_r},  \argminA_{R} \beta = \{R|\mu_r = \varepsilon_r\}$, where $\mu_r$ and $\varepsilon_r$ are the relative permittivity and permeability of the meta-atoms, respectively. Finally, the waveguiding function can be achieved by absorbing the incident energy and then pass along to adjacent meta-atoms through the waveguide layer as shown in Fig.~\ref{fig:meta_description}.

\vspace{-2mm}
\section{End-to-end System Characterization and Performance Evaluation}
\label{sec:e2e}
The realization of UM MIMO communications requires the development of accurate channel models able to capture the impact of both the plasmonic arrays as well as the behavior of a very large number of parallel waves propagating in space. In this section we explore these characteristics.

\vspace{-2mm}
\subsection{3D End-to-end Channel Model}
While we have studied the UM MIMO platforms in modes of transmission and reception~\cite{han2018ultra}, we herein extend the 3D end-to-end channel model to other aforementioned operation modes. Following a similar methodology, we can write the reflection coefficient introduced by the reflectarray at the frequency $f$ as
\vspace{-3mm}
\allowdisplaybreaks
\begin{equation}
\begin{aligned}
A(f)  &=  \sum_{n=0}^{N-1} \sum_{m=0}^{M-1} R_{mn}\left(f,\theta_t^{(m,n)},\phi_t^{(m,n)},\psi^{(m,n)} \right) \\
&\times\exp \left(\vphantom{\psi^{(m,n)} } jk_0 \left[x_{mn}\sin\left(\theta_t\right)\cos\left(\phi_t\right) \right. \right. \\
&\left.{}\left.{}+ y_{mn}\sin\left(\theta_t\right)\sin\left(\phi_t\right) \right] +j\psi^{(m,n)} \right),
\label{eq:reflectarray_gain}
\end{aligned}
\vspace{-2mm}
\end{equation}
where $M, N$ are the number of cells in the x-axis and y-axis of the integrated transmit--receive arrays which include both the UM MIMO array at the transceivers and the reflectarrays in the propagation environment.
$R_{mn}$ denotes the magnitude of the reflection coefficient of the $(m,n)^{\mathrm{th}}$ cell of the integrated array, which depends on the plasmonic element properties as well as the angle of departure in the elevation and azimuth planes, denoted by $(\theta_t, \phi_t)$. 
In addition, $k_0 = 2\pi f/c$ is the wavenumber with $c$ being the speed of light, and $x_{mn}$ and $y_{mn}$ represent the geometric location of the $(m,n)^{\mathrm{th}}$ cell.
Moreover, A phase shift $\psi^{(m,n)}$ is introduced between the incident and the reflected field from the $(m,n)^{\mathrm{th}}$ integrated array cell, to allow the phase of the reflected field to be uniform in a plane normal to the direction of the beam. 

In the propagation channel, a very large number of waves (or paths) is emitted from the transmitter with plasmonic arrays and will interact with the propagation environment until reaching the receiver. Existing channel modeling approaches normally treat the environments as passive channels, which equalization has to be performed at the receiver to combat the inter-symbol interference (ISI). However, with the utilization of the UM MIMO platforms, the plasmonic reflectarrays can manipulate the paths in the environments to enhance signal transmissions and diminish the effect of ISI. Based on the types of interaction, we characterize the channel as follows,
\vspace{-3mm}
\allowdisplaybreaks
\begin{align}
h(t) = & \alpha_{\mathrm{LoS}}\delta(t - \tau_{\mathrm{LoS}})\mathds{1}_{\mathrm{LoS}} + \sum_{s = 1}^{S}\alpha_{\mathrm{abs}}\delta(t - \tau_{\mathrm{abs}}) \nonumber \\
&+\sum_{i=1}^{I}\sum_{j = 1}^{J}\left( \alpha_{\mathrm{ref}}^{(i,j)} \prod_{m = 0}^{J-1}g_{m}^{(i,j)}\right ) \delta(t - \tau_{\mathrm{ref}}^{(i,j)}) \nonumber \\
&  +\sum_{w = 1}^{W}\alpha_{\mathrm{wvg}}^{(w)}g_m\delta(t - \tau_{\mathrm{wvg}}),
\label{eq:mpc_channel}
\vspace{-2mm}
\end{align} in which $\alpha_{(\cdot)}$ denotes the amplitude of paths in different types, including line-of-sight (LoS), absorption (abs), reflection (ref), and waveguiding (wvg). $\tau_{(\cdot)}$ is the time of the interaction between path and environment, $\mathds{1}_{LoS}$ is the presence indicator function of the LoS path, $i$ and $j$ denote the indices of reflection paths and segment in each path, $g_m^{(i,j)}$ is the loss induced by the $m$-th meta-atom upon wave impinging for the $j$-th path segment in the $i$-th path. The waveguided path only experiences one loss from the impinging meta-atom, then the path can be losslessly transmitted to the destination through the inter-connection among meta-atoms.
\vspace{-2mm}

\begin{figure*}[!t]
\vspace{-6mm}
	\centering
	\subfigure[Contour plot of average received power in the baseline case (no UM MIMO platforms).]{ \includegraphics[width=.28\textwidth]{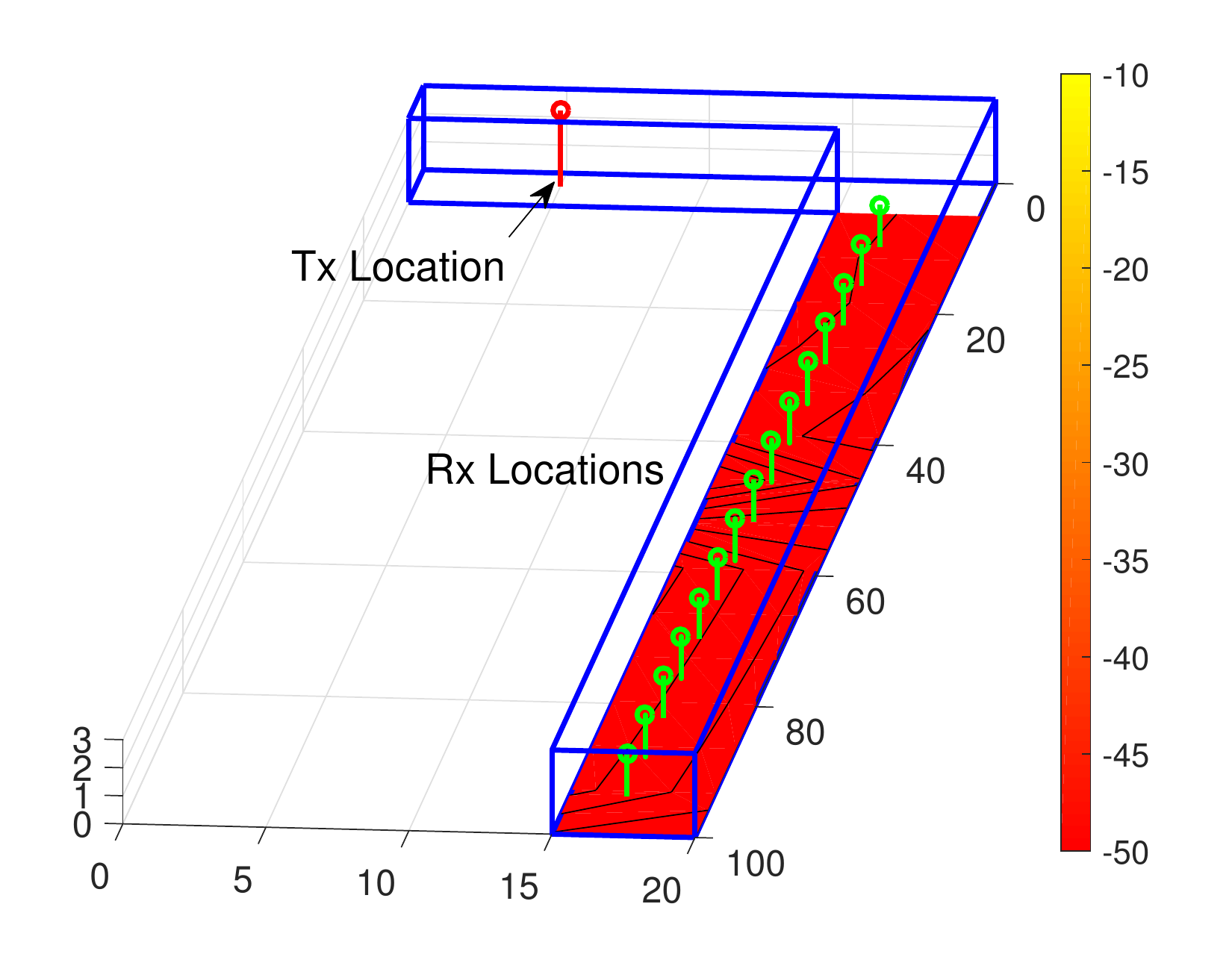} \label{fig:baseline}}
	\qquad
	\subfigure[Contour plot of average received power with UM MIMO platforms.]{\includegraphics[width=.28\textwidth]{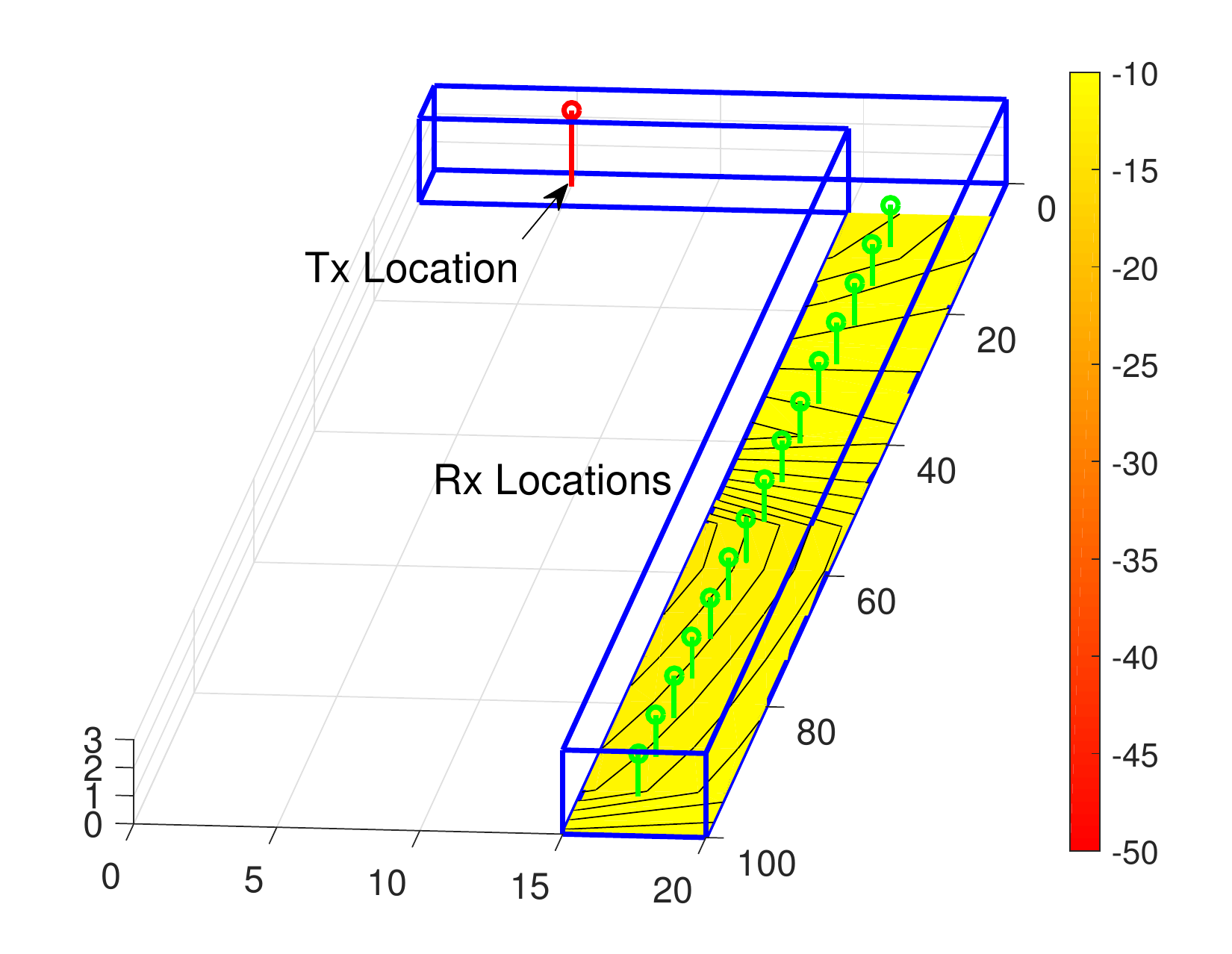} \label{fig:improve}}
	\qquad
	\subfigure[Achievable data rates in UM MIMO platforms-equipped environment at $300~GHz$.]{\includegraphics[width=.33\textwidth]{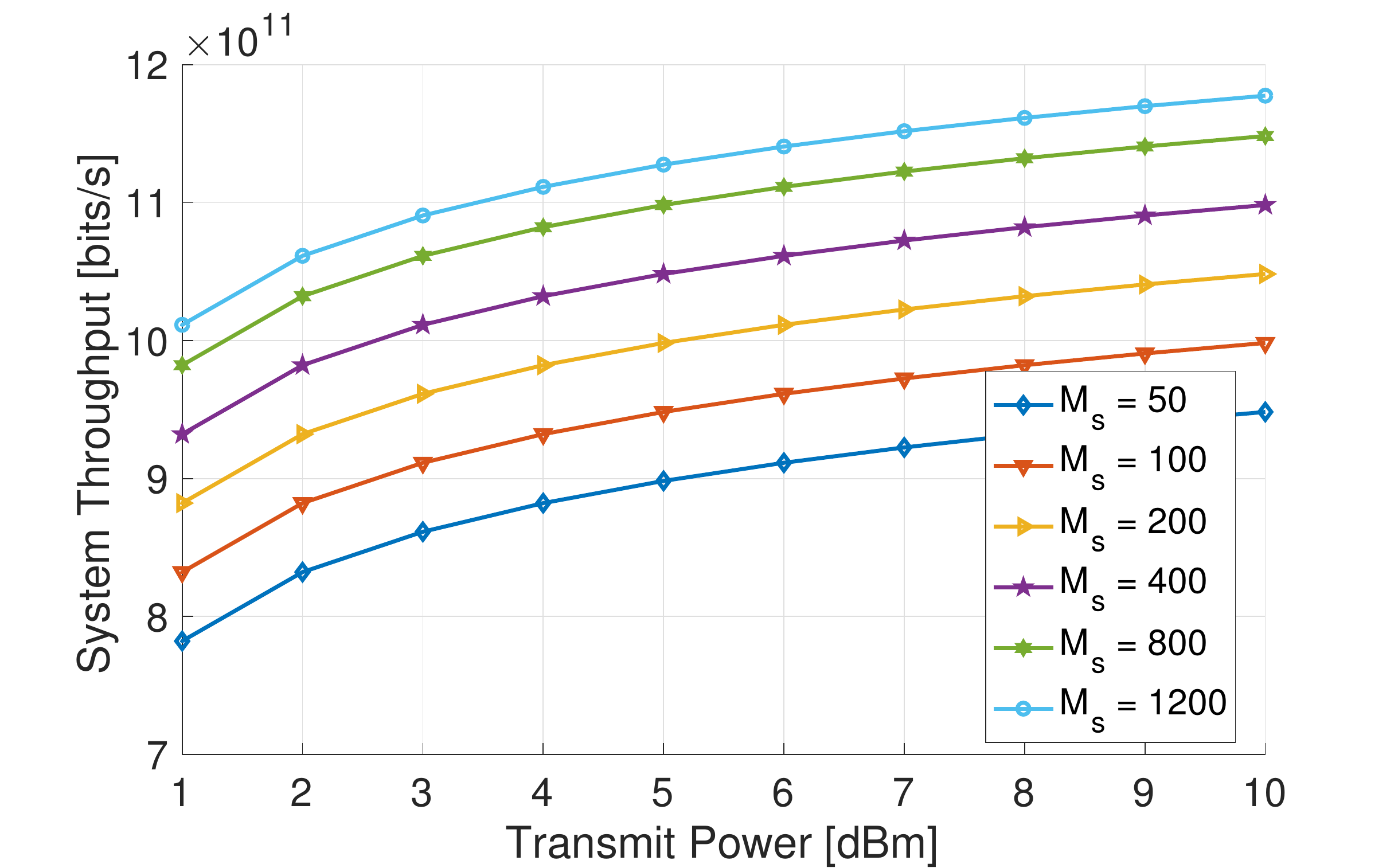} \label{fig:throughput}}
	\caption{Improvement in communication distance and system throughput under the deployment of UM MIMO platforms.}
	\label{fig:simulation}
	\vspace{-2mm}
\end{figure*}
\vspace{-2mm}
\subsection{Optimization in Resource Allocation}
Based on the above multipath channel model, the subsequent step is finding the desirable resource allocation strategies enabled by the UM MIMO platforms in transmission, reception, reflection, and waveguiding. This can be formulated as an optimization problem aimed at finding the optimal transmit power, $P_t^{(k)}$, the number of antennas in the UM MIMO systems in transmission and reception, $N_a^{(k)}$, and the number of antennas in the reflectarrays, $M_s^{(k)}$, for the  $k^\mathrm{th}$ receiver in the mm-wave and THz-band network with a total of  $K$ users. 
The objectives are to maximize the sum of product of the transmission distance $d_k$ and the overall throughput $\Gamma_k$, as follows:
\vspace{-2mm}
\allowdisplaybreaks
\begin{align}
\textit{Given: } &\mathbf{T}, \mathbf{R}, y_r^{(k)}, z_r^{(k)})^\intercal, P_t^\mathrm{tot}, N_a^\mathrm{tot}, M_s^\mathrm{tot} \\
\textit{Find: } & P_t^{(k)}, N_a^{(k)}, M_s^{(k)} \label{find}\\
\textit{Objective: } &\max  \sum d_k, \max \sum \Gamma_k\label{objective}\\
\textit{Subject to: } 
 & \sum P_t^{(k)} \leq P_t^\mathrm{tot} , \sum N_a^{(k)} \leq N_a^\mathrm{tot}, \\
& \Gamma_k\left(P_t^{(k)}, N_a^{(k)}, M_s^{(k)}\right) \geq \Gamma_\mathrm{th}^{(k)},\\
& \sum M_s^{(k)} \leq M_s^\mathrm{tot}, \text{  for  }k \in K. 
\vspace{-12mm}
\end{align}
In this optimization problem, $\mathbf{T}$ and $\mathbf{R}$ denote the 3D coordinates of the transmitter and the $k^\mathrm{th}$ receiver. The resources include the transmit power, number of antennas in the UM MIMO, and number of elements in the reflectarrays upper-bounded by $P_t^\mathrm{tot}$, $N_a^\mathrm{tot}$, and $M_s^\mathrm{tot}$, respectively. Moreover, the data rate, depending on the receiver's location, transmit power, number of allocated antennas in the integrated UM MIMO platform, needs to satisfy the threshold $\Gamma_\mathrm{th}^{(k)}$.

\vspace{-4mm}
\allowdisplaybreaks
\subsection{Performance Evaluation}
\label{sec:eval}
Based on the resource allocation optimization scheme, we can evaluate the performance of the UM MIMO platforms analytically in an end-to-end channel, the following metrics are taken into account: the achievable distance with satisfying signal-to-noise-ratio (SNR) and the maximum system throughput.  We start by considering a channel with the assumption of \textit{favorable propagation}~\cite{ngo2014aspects} between a BS with UM MIMO system with $N_a$ antenna and plasmonic reflectarray with the number of antennas $M_s$ for a user where the channel coefficient can be expressed as $h_{n, m} = g_{n, m}\sqrt{\alpha_n}$, where $g_{n,m}$ denotes the small-scale fading coefficient assumed to be  i.i.d. random variables with zero mean and unit variance for the pair of $n$-th antenna at BS and $m$-th antenna element in the plasmonic reflectarray and $\alpha_n$ is the large-scale fading coefficient taking the magnitude of Eq.~(\ref{eq:mpc_channel}). The channel transfer matrix seen by the user can be expressed as
\begin{equation}
\mathbf{H} = \left(\begin{smallmatrix}
h_{1,1} & \cdots  & h_{n,1}\\ 
 \vdots & \ddots  & \vdots \\ 
 h_{1,m}&  \cdots & h_{n,m}
\end{smallmatrix}\right) = \mathbf{GA}^{1/2}, \mbox{ with} 
\end{equation}
\begin{equation}
\mathbf{G} = \left(\begin{smallmatrix}
g_{1,1} & \cdots  & g_{n,1}\\ 
 \vdots & \ddots  & \vdots \\ 
 g_{1,m}&  \cdots & g_{n,m}
\end{smallmatrix}\right), \mathbf{A} = \textup{diag}\left(\alpha_1, \cdots , \alpha_n\right).
\end{equation}

In a downlink channel, the received signal vector $\mathbf{y}$ can be expressed as a function of transmitted signal vector $\mathbf{x}$, downlink channel matrix $\mathbf{H}$, and  complex Gaussian noise with zero-mean and unit-variance $\mathbf{C}\boldsymbol{N}$ as $\mathbf{y} = \sqrt{P_t}\mathbf{Hx} + \mathbf{C}\boldsymbol{N}$. Based on the assumption that the given resource optimization, the UM MIMO platforms have perfect knowledge of the channel, the achievable system throughput is
\begin{align}
C = B \log_2 \det(\mathbf{I} + P_t\mathbf{H}^H\mathbf{H}) = B \log_2(1 + P_tMN),
\end{align} 
which is upper-bounded by the size of the antenna arrays in the plasmonic transmit-receive arrays and the plasmonic reflectarrays, when the number of users are fewer in the same environment. Therefore, interference is asymptotically zero in this case. Fig.~\ref{fig:simulation} shows the distance enhancement and achievable data rates with different sizes in UM MIMO platforms at a center frequency $f_c = 300$~GHz with a bandwidth of $50$~GHz. The comparison between the average received power in an ``L''-shaped indoor corridor depicted in Fig.~\ref{fig:baseline} and Fig.~\ref{fig:improve} shows that the UM MIMO platforms can help to extend the coverage by around 40~dB. By utilizing a size of $N_a = 1024\times1024$ plasmonic transmit-receive arrays and with various numbers of units $M_s$ in our optimization function Eq.~(\ref{find}) (ranging from 50 to 1200) of plasmonic reflectarrays in the environment, the sum data-rate achievable in that environment with UM MIMO platforms can reach the Tbps level, as shown in Fig.~\ref{fig:throughput}.

\vspace{-4mm}
\section{Conclusion}
\label{sec:conclusion}
In this paper, UM MIMO platforms consisting of plasmonic antenna arrays, enabled by graphene and metasurfaces and utilized in transmission, reception, reflection and waveguiding, have been proposed. An end-to-end analytical physical model has been developed and exploited to highlight the significant enhancement in both transmission distance and achievable data-rate. Our future work includes leveraging the UM MIMO platforms with artificial intelligence potentials to further optimize the system performance.

%




\balance
\bibliographystyle{IEEEtran}
{\footnotesize
\bibliography{References}}

\end{document}